\documentstyle[12pt,agums]{article}

\lefthead{AFRAIMOVICH ET AL.}

\righthead{IONOSPHERIC RESPONSE TO THE TOTAL SOLAR ECLIPSE OF
JUNE 21, 2001}

\received{} \revised{} \accepted{} \paperid{ISZF0201}

   \cpright{PD}{2002}

\ccc{0148-0227/98/98JA-00000\$09.00}

\authoraddr{E. L. Afraimovich, O. S. Lesyuta,
Institute of Solar-Terrestrial Physics SD RAS, P.~O.~Box~4026,
Irkutsk 664033, Russia. (afra@iszf.irk.ru; leos@iszf.irk.ru)}
\slugcomment{To appear in the {\it ru.arxiv.org}, 2002.}

\begin{document}

%
%

\title{Ionospheric response to the total solar eclipse
of June 21, 2001 as deduced from the data from the African GPS
network}

\author{E. L. Afraimovich, O. S. Lesyuta}

\affil{Institute of Solar-Terrestrial Physics SD RAS, Russia}

\begin{abstract}
We discuss the measurements of the main parameters of the
ionospheric response to the total solar eclipse of June 21, 2001.
This study is based on using the data from three stations of the
global GPS network located in the area of the totality band in
Africa. This period was characterized by a low level of
geomagnetic disturbance (the Dst-index varied from -6 to 22 nT),
which alleviated greatly the problem of detecting the ionospheric
response to eclipse. An analysis revealed a clearly pronounced
effect of a decrease (depression) of the total electron content
(TEC) for all GPS stations. The delay between the smallest value
of the TEC with respect to eclipse totality was 9-37 min. The
depth and duration of the TEC depression were 0{.}5-0{.}9 TECU and
30-67 min, respectively. The results obtained in this study are in
good agreement with earlier measurements and theoretical
estimations.
\end{abstract}

\begin{article}

\section{Introduction}

Experimental observations of the ionosphere at the time of solar
eclipses provide a source of information about the character of
behavior of  the various  ionospheric parameters. Regular
ionospheric  effects of solar eclipses are fairly well understood.
They imply an increase of effective reflection heights,  a
reduction in concentration in the F-layer maximum, and a decrease
in total electron content (TEC) in the ionosphere,  which is
typical of the transition to the nightside ionosphere
[\markcite{{\it Cohen,} 1984}]. The behavior of  the above
parameters can  be modeled using appropriate ionospheric models
[\markcite{{\it Boitman et al.,} 1999;} \markcite{{\it Stubbe,}
1970}].

The main parameters of the ionospheric response include the value
of the delay  $\tau$ with respect to the eclipse totality phase,
as well as its amplitude $A$ and duration $\Delta T$. Almost all
publications devoted to the study of the ionospheric response to
solar eclipses make estimates of these parameters.  A knowledge of
these values makes  it possible  to refine, in terms of the
respective aeronomic ionospheric models, the time constants of
ionization, and the recombination processes at different heights
in the ionosphere.

The statistic of measurements of these parameters according to
published data is presented in Table~1 (columns 5, 6 and 7,
respectively). Table~1 also includes: Date - date of the total
solar eclipse; Location - geographic region where the eclipse was
observed or the path for the methods recording the delay time on
the VLF signal ray path between the signal reception site and the
transmitting station (VLF); FDS - frequency Doppler shift on the
HF ray path between the signal reception site and the transmitting
stations, as well as for the method of oblique-incidence
ionospheric sounding (OIS); Method - method used in the
investigation; Reference - reference to publication. The Note
column (column 9 of Table~1) presents the time resolution of the
method used in investigating the ionospheric response to total
solar eclipse, and the number of stations. The following
abbreviations are used in Table~1: I - vertical-incidence
ionospheric sounding; DS - differential Doppler shift; GPS -
Global Positioning System; $N_{m}F2$ - electron density in the
F2-layer maximum; $h_{m}F2$ - height of the F2-layer maximum;
$f_{0}F2$ - F2-layer critical frequency; $f_{0}E$ - E-layer
critical frequency; $h_{p}F2$ - virtual height of the F2-layer;
$h'F2$ - virtual height of the lower boundary of the F2-layer;
$h'$ - virtual height at fixed plasma frequencies; $f_{D}$ -
frequency Doppler shift; $A$ - signal amplitude; and St - number
of stations. The TECU (Total Electron Content Units), which is
equal to $10^{16}$ $m^{-2}$ and is commonly accepted in the
literature, will be used throughout the text.

Measurements  of $\tau$ were made by analyzing the characteristics
of the ionosphere-reflected radio signal at vertical-incidence
soundings  at a  network  of  ionospheric stations [\markcite{{\it
Borisov et al.,} 2000;} \markcite{{\it Cheng et al.,} 1992;}
\markcite{{\it Datta et al.,} 1959;} \markcite{{\it Huang et al.,}
1999;} \markcite{{\it Walker et al.,} 1991}].
In the cited references, the value of $\tau$ was found to vary
from 5 min [\markcite{{\it Walker et al.,} 1991}] (line 2 of Table
1) to 80 min [\markcite{{\it Huang et al.,} 1999}] (line~7 of
Table~1) according to the $N_{m}F2$ data, and from 9.5 min
[\markcite{{\it Borisov,} 2000}] (line~4 of Table~1) to 30 min
[\markcite{{\it Walker et al.,} 1991}] according to the $f_{0}F2$
data. The amplitude $A$ of a decrease in local electron density
from 0.2 to 1$\times$$10^{12}$ m${}^{-3}$ ($N_{m}F2$), and from
0.2 to 1.4 MHz ($f_{0}F2$). The response duration $\Delta
T$=100-240 min according to the data on $N_{m}F2$ and $f_{0}F2$.

To analyze the ionospheric effects from the total solar eclipse of
September 23, 1987, \markcite{{\it Cheng et al.,} [1992]}  used
the phase variation of the VLF signal transmitted from NDT, Yosami
(34.97$^\circ$ N; 137.02$^\circ$ E), Japan, and recorded at
Kaojong (24.95$^\circ$ N; 121.15$^\circ$ E), Taiwan, as well as
the differential Doppler shift data from the Lunping Observatory
(25$^\circ$ N; 121.17$^\circ$ E). Results of this investigation
are presented in line~3 of Table~1.

Interesting results were obtained by investigating the ionospheric
response to the total solar eclipse of March 9, 1997
[\markcite{{\it Boitman et al.,} 1999}] (line~5 of Table~1) and of
August 11, 1999 [\markcite{{\it Cherkashin and Agafonnikov,}
2001}] (line~6 of Table~1) using FDS and OIS [\markcite{{\it
Boitman et al.,} 1999;} \markcite{{\it Borisov et al.,} 2000}]
 (lines~5 and 4 of Table~1).

\markcite{{\it Boitman et al.,} [1999]} used the following radio
sounding paths: Tory (51.7$^\circ$ N; 103.8$^\circ$ E)-Irkutsk,
Tory-Ulan-Ude, Tory-Krasnoyarsk, Tory-Chita (recording of FDS),
and Tory-Irkutsk (OIS method). To investigate the ionospheric
response to the total solar eclipse of March 9, 1997,
\markcite{{\it Borisov et al.,} [2000]} used frequency Doppler
shift data obtained from soundings for the following paths:
Novosibirsk-Tomsk, Krasnoyarsk-Tomsk, Magadan-Yakutsk,
Komsomolsk-on-Amur-Yakutsk, Khabarovsk-Yakutsk, Yakutsk-Tomsk,
Magadan-Tomsk, Irkutsk-Tomsk, and Thushima-Tomsk. According to the
data obtained using OIS, $\tau$ = 6-20 min, $A$ = 0{.}5-1{.}5 Hz
(for short paths) and 2{.}3-3 Hz (for long paths), and $\Delta
T$=40-150 min (lines~4 and 5 of Table~1).

\markcite{{\it Cherkashin and Agafonnikov,} [2001]} investigated
the ionospheric response to the total solar eclipse of August 11,
1999 for the following paths: London (GB)-Troitsk (IZMIRAN) with
L=2.5 Mm (at 12095 kHz frequency), and Prague (Czech
Republic)-Troitsk (at 9520 kHz frequency). In the cited reference
the value of the frequency Doppler shift was converted to the
amplitude A of the variation of the reflection height (line~6 of
Table~1). The values of $\tau$ and $\Delta T$, obtained by
\markcite{{\it Boitman et al.,} [1999]} and \markcite{{\it
Cherkashin and Agafonnikov,} [2001]} using FDS are 0-15 min and
87-120 min (lines~5 and 6 of Table~1), respectively.

The development of the global navigation system GPS and the
creation, on its basis, of extensive networks of GPS stations
(which at  the  end  of 2001 consisted of no less than 900 sites),
the data from which are placed on the Internet [\markcite{{\it
Klobuchar,} 1997}], opens up a new era  in  remote sensing  of the
ionosphere.  At almost any point of the globe and at any time at
two coherently-coupled frequencies $f_1=1575{.}42$ MHz and
$f_2=1227{.}60$  MHz, two-frequency multichannel receivers of the
GPS system are used to carry out high-precision measurements of
the group and phase delay along the line of sight between the
ground-based receiver and satellite-borne transmitters in the zone
of  reception.  The sensitivity afforded  by  phase measurements
in the GPS system permits irregularities to be detected with an
amplitude of up  to $10^{-3}$ of the diurnal variation of TEC
[\markcite{{\it Melbourne et al.,} 1994}].

A large body of data of analysis of the ionospheric response to
total solar eclipse was obtained using the GPS [\markcite{{\it
Afraimovich et al.,} 1998;} \markcite{{\it Afraimovich et al.,}
2001;} \markcite{{\it Feltens,} 2000;} \markcite{{\it Huang et
al.,} 1999;} \markcite{{\it Tsai and Liu,} 1999}].
The values of $\tau$, $A$ and $\Delta T$, derived from
investigating the ionospheric response to total solar eclipses
using the GPS, differ greatly ($\tau$ varies from 4 to 16 min, $A$
ranges from 0{.}1 to 3 TECU, and $\Delta T$ varies between 32 and
75 min). The large scatter of the values of $\tau$, $A$ and
$\Delta T$ is likely to be associated with the difference of the
longitude and latitude ranges, over which the investigations were
carried out, the season, the technique for processing the GPS
data, as well as with differing geomagnetic situations.

To investigate the ionospheric response to the total solar
eclipses of March 9, 1997 and August 11, 1999, \markcite{{\it
Afraimovich et al.,} [1998, 2001]} used the variations in total
"oblique" electron reflection and the "vertical" TEC value,
respectively. The study of the ionospheric response to the August
11, 1999 total solar eclipse is based on using the data from about
100 GPS stations located in the neighbourhood of the eclipse
totality phase in Europe.

In examining the ionospheric effect from the total solar eclipses
of October 24, 1995 and March 9, 1997, \markcite{{\it Huang et
al.,} [1999]} and \markcite{{\it Tsai and Liu,} [1999]} analyzed
the variations in "vertical" TEC. \markcite{{\it Tsai and Liu,}
[1999]} carried out their investigations near the magnetic
equator; for that reason, the TEC series contain, in addition to
the response to total solar eclipse, variations caused by the
dynamics of the equatorial anomaly.

\markcite{{\it Feltens,} [2000]} investigated the ionospheric
response to the August 11, 1999 total solar eclipse using total
electron content maps (TEC maps, line~10 of Table~1), the raw data
for which are represented by TEC series obtained  by means of the
global network of GPS receivers. Taking into account the temporal
resolution of standard TEC maps (2 hours) as well as the parameter
$\Delta T$ derived from analyzing earlier work (1-1{.}5 hour) it
can be concluded that such TEC maps do not secure the necessary
determination accuracy  of $\tau$, $A$ and $\Delta T$.

Hence a  large  body  of  experimental  data do not permit us to
make any reliable estimates of the basic parameters of the
ionospheric  response. One of the reasons for such a great
difference is that different methods of measurements are used,
which differ greatly by spatial  and temporal resolution. However,
the  main reason is caused by dissimilar characteristics of the
eclipse itself,  by  geophysical conditions  of individual
measurements, and  by  a  large difference in latitude, longitude
and local time when experiments are conducted.

To obtain more reliable information about the behaviour of the
ionosphere during an eclipse requires simultaneous measurements
over a large  area  covering regions with different local times.
Furthermore,  high spatial  (of some  tens  of kilometers at
least) and temporal (at least 1 min) resolution is  needed.
However,  none  of  the above familiar methods meets such
requirements.

In this paper the method of detecting the ionospheric response to
solar eclipse, reported in \markcite{{\it Afraimovich et al.}
[2001]}, is used to estimate the main parameters of the
ionospheric response to the total solar eclipse of June 21, 2001.

\section{The geometry and general  information  of  total  solar  eclipse
June 21, 2001}

The total solar eclipse of June 21, 2001 began in the South
Atlantic Ocean 400 km to the south of Uruguay. The Moon's shadow
made its first contact with the surface at 10{:}35 UT. The eclipse
duration at that point was 2 min 6 s. Within two subsequent hours
the shadow was moving along the surface of the South Atlantic
Ocean. Totality occurred at 12{:}04 UT at the point with the
coordinates $11.25^\circ S$; $2.75^\circ E$ and lasted for 4 min
56 s. The velocity of the Moon's shadow along the surface
averaged 1{.}2 km/s. For 2 h 56 min (duration of the total solar
eclipse), the Moon's shadow traveled a distance of about 12000 km.

Figure 1 presents the schematic map of the movement of the Moon's
shadow along the surface in Africa (based on using the data from
\markcite{{\it Espenak and Anderson} [2001]}). The heavy line
shows the centerline of the eclipse at the ground level, and thin
line show the southern and northern boundaries. The location of
the reference ionospheric station Madimbo ($22{.}38^\circ S$;
$30{.}88^\circ E$) is shown by the heavy cross symbol. Heavy dots
correspond to the locations of the GPS stations used in the
analysis; their geographic coordinates are presented in Table 1.
Numbers for the longitudes of $10^\circ$, $20^\circ$, $30^\circ$,
$40^\circ$ and $50^\circ E$ correspond to the local time of
eclipse totality at ground level.

In this paper we have confined ourselves only to analyzing the
area from the western coast of Africa to the point with the
coordinates $19{.}95^\circ S$; $40{.}07^\circ E$, where the phase
of totality was observed at 13{:}25 UT (16{:}09 LT). Thus the
eclipse effect in the region under investigation took place for
the conditions of the dayside summer ionosphere.

Fig.~2d shows the Dst-variations for June 20 (solid curve), June
20 (solid curve with dots), and June 22 (solid curve with
triangles) 2001. This period was characterized by a low level of
geomagnetic disturbance (within -6 - 22 nT), which alleviated
greatly the problem of detecting the ionospheric response to
eclipse.

\section{The ionospheric response by eclipse  from  date  of  ionospheric
station Madimbo}

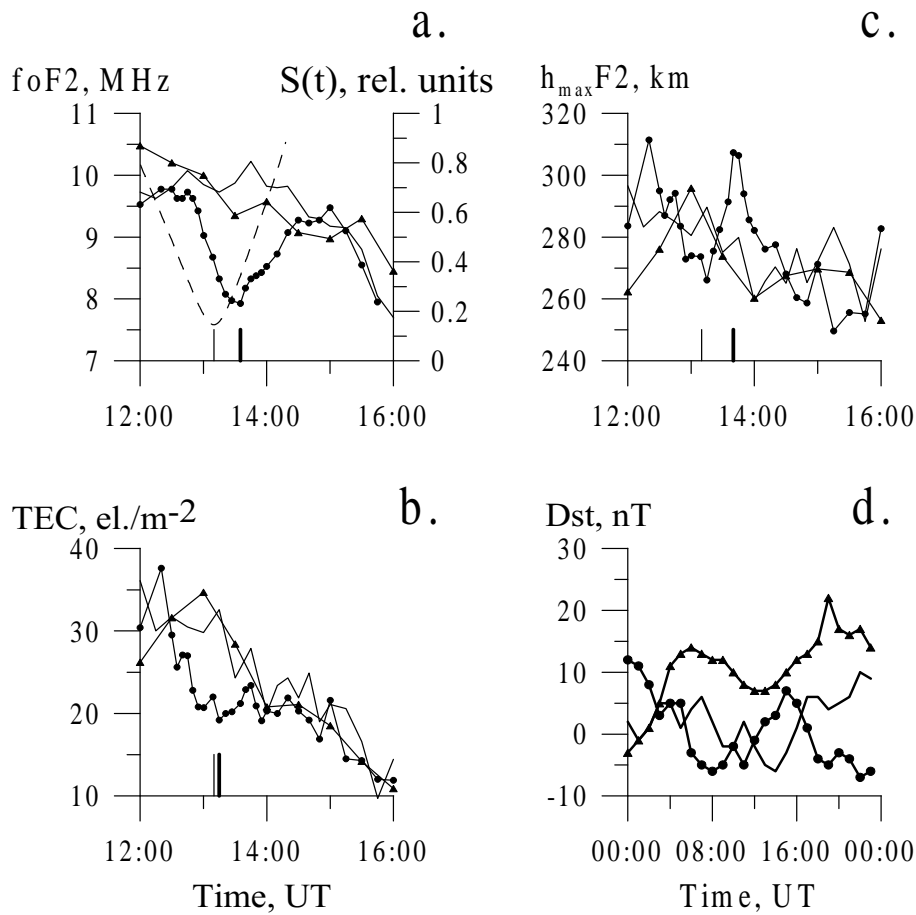
First we consider the variations of ionospheric parameters over
the time interval 12{:}00 - 16{:}00 UT on June 21 and on the
background days of June 20 and 22, 2001, using the data from
station Madimbo - Fig.~2a, b, c. Solid curves with dots correspond
to the variations in critical frequencies $f_{0}F2$ (panel a), TEC
(panel b), and the heights of the F2-layer maximum ($h_{m}F2$) for
June 21, 2001 (from here on the heights of the F-region and higher
altitudes are implied). Solid curves and solid curves with
triangles show the variations of the parameters $f_{0}F2$, TEC,
and $h_{m}F2$ for June 20 and 22, 2001, respectively. The dashed
curve on panel a shows the geometrical function of eclipse S(t)
obtained for ionospheric station Madimbo during 12-14{:}22 UT.The
geometrical function of eclipse $S(t)$ represents a part of the
solar disk area that is not occulted by the Moon's shadow, and is
expressed in fractions of this area in arbitrary units. To
calculate the geometrical function of eclipse, we developed a
special program which can be used to calculate S(t) at any
heights. The mathematical apparatus that was used in preparing
this program, was described in a large number of publications
[e{.} g{.}, \markcite{{\it  Mikhailov,} 1954]}.

The onset time of the phase of totality (13{:}10 UT) at 300 km
level at the station's location is shown by a thin vertical line
(Fig.~2a).

The eclipse effect is most clearly distinguished in the
variations of critical frequencies $f_{0}F2$, the greatest
difference of which from background values on June 20 and 22 at
the time of a minimum (13{:}35 UT, thick vertical line in
Fig.~2a) was 1{.}5-2 MHz.  The delay between the times of
totality and a minimum $f_{0}F2$ was 25 min (the time resolution
of the data from the ionospheric data used in this study, is 5-15
min).

The delay $\tau$ between the times of a minimum $f_{0}F2$ and a
minimum of the function S(t) for ionospheric station Chilton
during the total solar eclipse of August 11, 1999 was 4 min
[\markcite{{\it Afraimovich et al.,} 2001}] (time resolution 4
min).

The difference of $\tau$ for the total solar eclipses of June 21,
2001 and August 11, 1999 seems to be caused by the different time
resolution used in the analysis of the $f_{0}F2$, as well as by
the different distance of ionospheric stations in latitude from
the band of totality.

The eclipse effect on other parameters (the TEC value obtained
using the data from ionospheric station Madimbo to the height of
the $f_{0}F2$ layer maximum and using the ionospheric model above
this maximum, as well as the height of the $h_{m}F2$ - Fig.~2b, c)
is not as clearly pronounced as in the case of $f_{0}F2$.
According to the TEC data, $\tau$ = 5 min, and for $h_{m}F2$ we
have $\tau$ = 30 min.

Similar results (for $f_{0}F2$) of measurements at the ionospheric
station were obtained during the solar eclipse of September 23,
1987 in South-East Asia [\markcite{{\it Cheng et al.,} 1992}].

\section{The process  of  GPS--network  data  and  results of analysis of
ionospheric  effect  by  total  solar  eclipse  of June 21, 2001}

We now describe briefly the sequence of procedures of processing
the GPS data. Primary data include series of the "oblique" value
of TEC $I(t)$, as well as the corresponding series of values of
the elevation $\theta(t)$ measured from the ground, and of the
azimuth $\alpha(t)$ of the LOS to the satellite measured eastward
from north. These parameters are calculated using our developed
program CONVTEC which transforms RINEX-files [\markcite{{\it
Gurtner,} 1993]}, standard for the GPS system, received via the
Internet. The series of values of the elevation $\theta(t)$ and
azimuth $\alpha(t)$ of the LOS to the satellite are used to
determine the location of subionospheric points. In the case under
consideration these results were obtained for elevations
$\theta(t)$ larger than $45^\circ$.

The snowflake symbol in Fig.~1 shows the location of the
subionospheric point at the time of a maximum TEC response.

Variations of the "oblique" TEC $I(t)$ are determined on the basis
of phase measurements at each of the spatially separated
two-frequency GPS receivers using the formula from [\markcite{{\it
Afraimovich et al.,} 1998]}:

\begin{equation}
\label{TSE-eq-01}
I_{0}=\frac{1}{40{.}308}\frac{f^2_1f^2_2}{f^2_1-f^2_2}
                           [(L_1\lambda_1-L_2\lambda_2)+const+nL]
\end{equation}

where $L_1\lambda_1$ and $L_2\lambda_2$ are additional paths of
the radio signal caused by the phase delay in the
ionosphere,~(m);  $L_1$ and $L_2$ represent the number of phase
rotations  at  the  frequencies  $f_1$  and $f_2$;
$\lambda_1$   and   $\lambda_2$   stand  for  the  corresponding
wavelengths,~(m);  $const$ is the unknown initial  phase
ambiguity,~(m); and $nL$~ are errors in determining the phase
path,~(m).

The series of values of the elevation $\theta(t)$ and azimuth
$\alpha(t)$ of the LOS to the satellite were used to determine the
location of subionospheric points and to transform the "oblique"
TEC $I_0(t)$ to the corresponding value of the "vertical" TEC
using the technique described in [\markcite{{\it Klobuchar,}
1986]}:

\begin{equation}
\label{TSE-eq-02} I = I_0 \times cos
\left[arcsin\left(\frac{R_z}{R_z + h_{m}}cos\theta\right) \right]
\end{equation}

where $R_z$ is the Earth's radius, and $h_{m}=300$ is the height
of the equivalent thin shell.

With the purpose of eliminating variations of the regular
ionosphere, as well as trends introduced  by  the  satellite's
motion, we  employ  the procedure of eliminating the trend by
preliminarily smoothing the initial series with the time window in
the range from 40  to  60  min  which  is fitted  for  each TEC
sampling. The selection of the time window from 40 to 60 min when
removing the trend is reconciled with the expected duration of the
ionospheric response. The determination accuracy of the time of
the TEC response extremum at the chosen values of the time window
was sufficiently high (not worse than 1 min). The reason is that
under the conditions of the magnetically quiet day of June 21,
2001 (the largest value of Dst did not exceed 12 nT) the amplitude
of TEC background variations was far below the amplitude of the
TEC response to the eclipse.

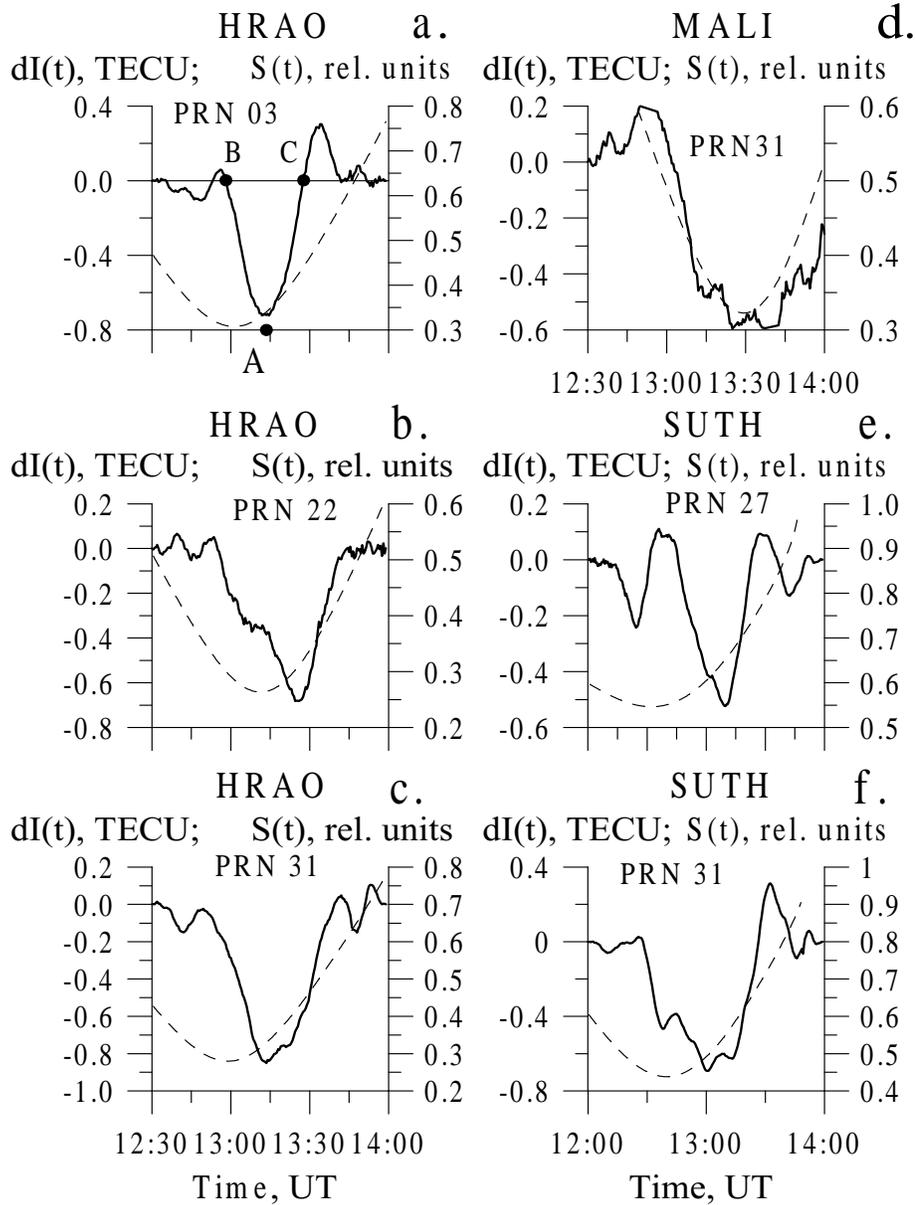
Fig.~3 presents the filtered variations of TEC $dI(t)$ for HRAO
stations for satellite N03 (PRN03) - Fig.~3a; HRAO (PRN22) -
Fig.~3b; HRAO (PRN31 - Fig.~3c; MALI (PRN31) - Fig.~3d; SUTH
(PRN27) - Fig.~3e; and SUTH (PRN31) - Fig.~3f for June 21, 2001
(heavy lines). These panels also show the geometrical eclipse
functions $S(t)$ at 300  km altitude calculated for the
corresponding subionospheric points. Onset times of minima
$t_{min}$ of the $dI(t)$ series are presented in Table~1. In
Table~1 the following symbols are used: PRN - satellite number;
SLAT and SLON - latitude and longitude of the subionospheric
point, respectively; $A$ - minimum value of $dI(t)$-variations;
$\Delta  T$ - response duration; and $\tau$ - delay between
minima of the $dI(t)$ and $S(t)$ series.

It is evident from Fig.~3a, the filtered variations resemble in
their form a triangle whose vertex (point A) corresponds to the
time, at which a minimum TEC value is attained. The value of
$dI_{min}$ itself can serve as an estimate of the amplitude of the
TEC response to eclipse, and the time interval between the times
of intersection of the $dI$=0 line (points B and C) can serve as
an estimate of the response duration $\Delta T$.

Such $dI(t)$ variations are characteristic for all the GPS
stations and satellite numbers 1, 3, 13, 22, 27 and 31 listed in
Table~1. That the above-mentioned satellites were chosen for the
entire set of GPS station was dictated by the fact that for these
satellites a maximum value of the elevation $\theta$ of the LOS
to the satellite, for the time interval 13{:}00-14{:}00~UT,
exceeded $45^\circ$, which minimized the possible error of
transformation to the "vertical" TEC value as a consequence of
sphericity.

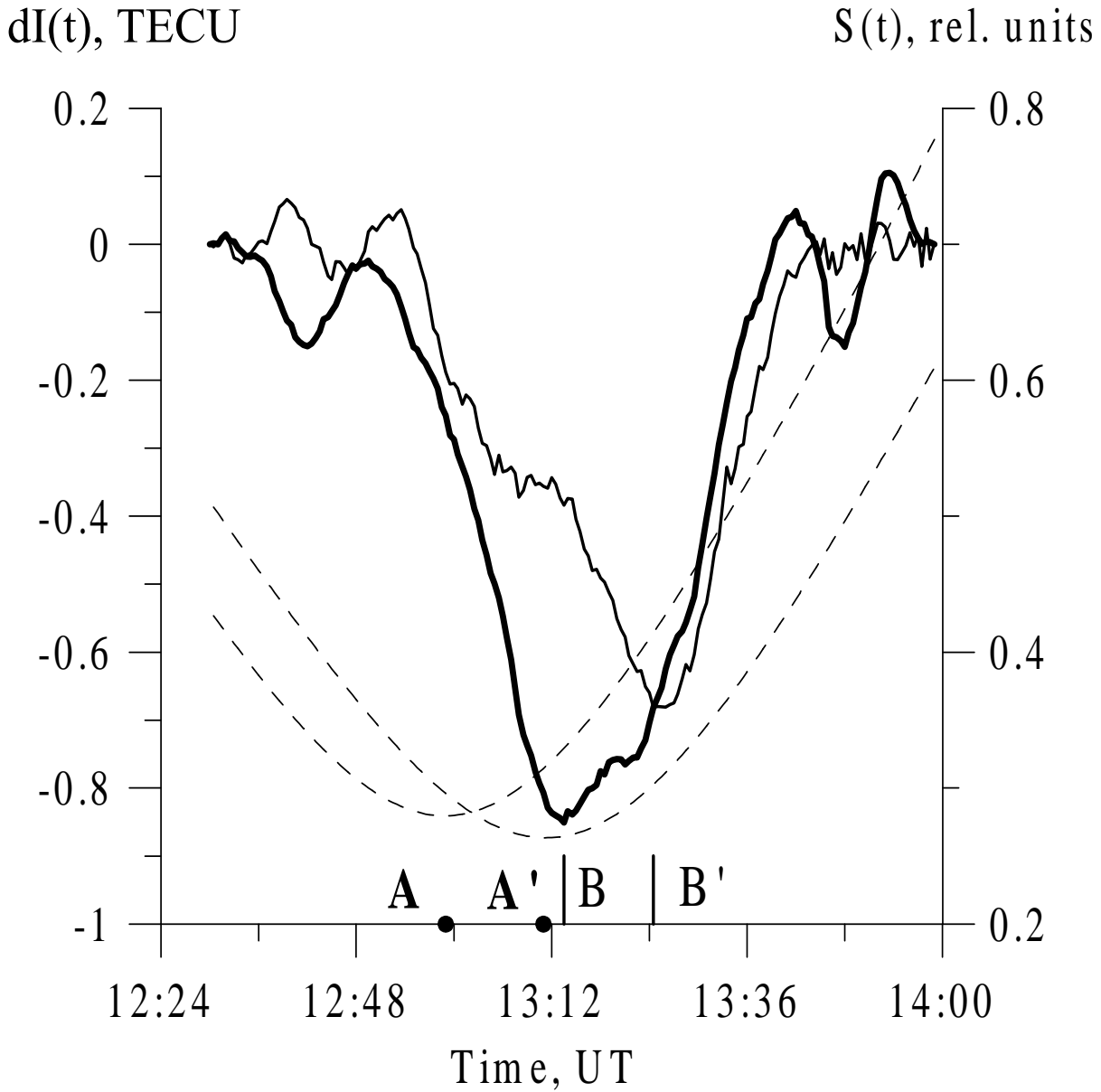
For comparison, Fig.~4 presents the $dI(t)$ - variations for
different subionospheric points of station HRAO: PRN22 - thin
solid line, and PRN31 - thick solid line separated by $7^\circ$ in
longitude. It is evident from the figure, the responses to eclipse
for the subionospheric points PRN22 and PRN31 are quite similar
both in their form and in amplitude, but the delay of the response
increases with longitude.

A thin dashed line and a thin dashed line with dots in Fig.~4
shows that $S(t)$ for HRAO (PRN22) and HRAO (PRN31), respectively.
The points $\bf A$ and $\bf A^{'}$ for GPS station HRAO (PRN31)
correspond to the time of totality and the time of a minimum of
the $dI(t)$, respectively. The points $\bf B$ and  $\bf B^{'}$ for
GPS station HRAO (PRN22) correspond to the time of totality and
the time of a minimum of the $dI(t)$, respectively. The use of
satellites PRN22 and PRN31 for the HRAO station was dictated by
the fact that the paths of these satellites lie in about the same
latitude range (Fig.~1). It is evident from Table 1 that the
longitudinal dependence of the time of a minimum $dI(t)$ is not a
monotonic function for all GPS stations in latitude from the
totality band, i{.} e{.} the geophysical conditions for some of
the trajectories differed drastically.

\section{Conclusion}

The results reported in this study are in good agreement with
earlier measurements and theoretical estimations (see a review of
the data in the Introduction). The principal difference of our
data is the higher reliability of determination of the main
eclipse response parameters caused by high space-time resolution
and by improved sensitivity of detection of ionospheric
disturbances using GPS.

To investigate the ionospheric response to the total solar
eclipses of March 9, 1997 and August 11, 1999, \markcite{{\it
Afraimovich et al.,} [1998, 2001]} used the variations in total
"oblique" electron reflection and the "vertical" TEC value,
respectively. The values of $\tau$, $A$ and $\Delta T$, derived
from investigating the ionospheric response to total solar
eclipses using the GPS, differ greatly ($\tau$ varies from 0 to
400 min, $A$ ranges from 0{.}1 to 15 TECU, and $\Delta T$ varies
between 32 and 200 min). The large scatter of the values of
$\tau$, $A$ and $\Delta T$ is likely to be associated with the
difference of the longitude and latitude ranges, over which the
investigations were carried out, the season, the technique for
processing the GPS data, as well as with differing geomagnetic
situations.

The TEC data for the total solar eclipses of August 11, 1999 and
June 21, 2001 were processes following the technique described in
the "The process  of  GPS--network  data  and results of analysis
of ionospheric  effect  by  total  solar eclipse  of June 21,
2001" section. The values of the parameters $\tau$, $A$ and
$\Delta T$, obtained by analyzing the ionospheric response to the
total solar eclipses of August 11, 1999 and June 21, 2001, differ
by no more than 5-21 min, 0{.}4-0{.}6 TECU, and 6-7 min,
respectively. The difference of the parameters $\tau$, $A$ and
$\Delta T$ can be explained by the difference of the geometry of
the eclipses as well as by the difference of the latitude and
longitude ranges.

The time constant of ionization decrease in the $F2$ maximum
exceeds greatly the duration of the totality phase of eclipse,
which leads to a decrease in the response amplitude. The TEC
response amplitude in terms of various models were made in
[\markcite{{\it Stubbe,} 1970}] and [\markcite{{\it Boitman et
al.,} 1999}], also for the spring season.

The value of $\tau$ for $f_{0}F2$ about 25 min corresponded to the
local time of 15{:}38~LT. The parameter $\tau$ for $f_{0}F2$ at
the time of the total solar eclipse of March 9, 1997 varied from
9{.}5 to 16 min according to the data reported by \markcite{{\it
Borisov et al., } [2000]}, which is in good agreement with our
data.

We cannot directly compare the physical significances of the delay
times $\tau$ obtained from $f_{0}F2$ observations with those
obtained from TEC measurements. $f_{0}F2$ refer to a given height,
namely the height of peak density. On the other hand, $\tau$,
which is height-dependent, gives a weighted value because $\tau$
is small at low heights, say, 200 km, and is large in the topside,
say, 1000 km. Thus $\tau$ will vary along the GPS-Ground ray path.
Since most of the TEC is above the peak the observed GPS-tau
should be larger than the $f_{0}F2$-$\tau$.

\markcite{{\it Ivelskaya et al., } [1977]}, using simulation
methods, showed that the variations of the delay time $\tau$ of a
minimum local electron density $N_e(t)$ with respect to a minimum
of the ion production function are: at 150 km altitude $\tau$ =
1-2 min, at 200 km - $\tau$ = 3 min, at 300 km - $\tau$ = 20 min,
and above 600 km - $\tau$ = 45 min. In this paper, $\tau$ was
estimated at about 3 min for 200 km altitude, and at 40 min for
300 km for 12 LT.

%

\acknowledgments We are grateful to Lee-Anne McKinnel for kindly
making the data from ionospheric station Madimbo available to us.
This work was done with support under RFBR grant of leading
scientific schools of the Russian Federation No.~00-15-98509 and
Russian Foundation for Basic Research (grants 99-05-64753,
00-05-72026, and 01-05-06171).

\end{article}
\end{document}